# Leveraging Smartphone Sensors for Detecting Abnormal Gait for Smart Wearable Mobile Technologies

Md. Shahriar Tasjid[1(✉)], Ahmed Al Marouf[1,2]
[1]Daffodil International University, Dhaka, Bangladesh
[2]University of Calgary, Alberta, Canada
shahriar15-9384@diu.edu.bd

**Abstract**—Walking is one of the most common modes of terrestrial locomotion for humans. When a person walks, there is a pattern in it, and it is known as gait. Gait analysis is used in sports and healthcare. We can analyze this gait in different ways, like using video captured by the surveillance cameras or depth image cameras in the lab environment. It also can be recognized by wearable sensors. e.g., accelerometer, force sensors, gyroscope, flexible goniometer, magneto resistive sensors, electromagnetic tracking system, force sensors, and electromyography (EMG). Analysis through these sensors required a lab condition, or users must wear these sensors. For detecting abnormality in gait action of a human, we need to incorporate the sensors separately. Understanding a regular gait vs. abnormal gait may give insights to the health condition of the subject using the smart wearable technologies. Therefore, in this paper, we proposed a way to analyze abnormal human gait through smartphone sensors. We can track down person's gait using sensors of these intelligent wearable devices. To do the stratification of the gait of the subjects, we have adopted five machine learning algorithms with addition a deep learning algorithm. The advantages of the traditional classification are analyzed and compared among themselves. After rigorous performance analysis we found support vector machine (SVM) showing 96% accuracy, highest among the tradition classifiers. 70%, 84%, and 95% accuracy is obtained by the logistic regression, Naïve Bayes, and k-Nearest Neighbor (kNN) classifiers, respectively. As per the state-of-the art, deep learning classifiers has been proven to outperform the traditional classifiers in similar binary classification problems. We have considered the scenario and applied the 2D convolutional neural network (2D-CNN) classification algorithm, which outperformed the other algorithms showing accuracy of 98%. The model can be optimized and can be integrated with the other sensors to be utilized in the mobile wearable devices.

**Keywords**—human gait detection, abnormal gait, machine learning, deep learning, sensors, accelerometer, gyroscope

## 1       Introduction

Human gait analysis is the study of assessing the functional aspects of human locomotion. It is used successfully various fields, including medicine, sports and

*Paper*—Leveraging Smartphone Sensors for Detecting Abnormal Gait for Smart Wearable Mobile…

ergonomics [1–3]. Through the analysis we can extract the pattern of walking of a person. Gait analysis can be used to assess the gait phase and kinetic parameters of human gait events and quantitative evaluation of musculoskeletal functions [4]. For that reason, Gait analysis is used in sports, health diagnosis, etc. In sports, the purpose of this type of analysis is to help athletes in their performance. Besides, it is used to treat individual in their walk or after injury recovery in the healthcare. Jawad et al. [5] studied the Ilizarov frame, which is used to treat complex fractures. They analyzed the spatiotemporal parameter of gait during Ilizarov technique treatment. This type of analysis usually requires lab environment, time, or labor-intensive engineering technologies, mainly on analysis where feature extraction is needed [6–8]. For this reason, machine learning algorithms are being fusing into gait and posture related investigations [9–11].

Gait analysis has been used successfully to detect risk of fall and prevent injuries from falling [12]. However, we can use this analysis for older people or sick people by detecting abnormal gait. Usually, sick people have different walking pattern. This can be recognized as abnormal gait. We can detect abnormal gait by using sensor from smartphone. This study proposed an accelerometer and gyroscope sensor to detect human gait as normal and abnormal. People who use a smartphone can use this easily with no need of extra assistance. One can integrate this work in the smartphone app. If abnormal gait is detected, then the system will notify the emergency contact person.

We have used one smartphone placed on hand to record the walking using accelerometer and gyroscope for this study. After recording the activities data is being processed and we have applied different classification algorithm on it to recognize normal and abnormal gait.

## 2 Related works

Machine learning algorithms are intensively used to detection in gait and posture nowadays. There are many works found in that they try to detect different walking along with other activities and posture. Luo et al. [13] they studied sensors data of Inertial Measurement Units (IMU) and the data analyzed to detect gait with machine learning. Kang et al. [14] they proposed an algorithm to detect motion of walking, as well as counting step with gyroscope data. Their work is good, and they have achieved an almost good accuracy of 95.74% with their proposed algorithm. They [15] proposed a real time method for gait event detection of electrically simulated paraplegic subjects. Takiguchi [16] studied the human the walking characteristics. They observe the electrical field created around the human body with walking. And They tried to detect walking with waveforms reflected on a living environment or a place where waveforms can be reflected. They [17] develops a sensing module worn on wrist and ankle of the subjects to collect motion signals and detected 10 daily activities. They applied NWFE algorithm and got an accuracy of 90.5% which is also a good work. They used accelerometer on user's wrist, chest and ankle for identify twelve activities. They [18] used "K nearest neighbor (KNN), rotation forest and backpropagation neural network (BPNN)" to identify those activities. They got an average accuracy of 98%. Gu et al [19] studied seven activities with smartphone sensors on independent motion state and applied least squares support vector machine (LS-SVM) classifier and got and accuracy of 90.7%. All this work has been done in a lab environment and instructions has been given to the subjects how to use the device. Although in this study we collected data putting smartphone in the subject's pocket and data is collected in outdoor in normal situation. They we combined and classified the data. They [20] monitored activities on patients and analyzed the data of accelerometer on patient's chest. They applied K nearest neighbor (KNN) and CART to identify those activities. They [21] used smartphone to recognize human activity. They used accelerometer and gyroscope data and used a deep learning approach which is Stacked Autoencoder (SAE) and they achieved accuracy of 97.5%. Chan et al. [22] studied gait assessment with low back pain and used accelerometer for data collection. They applied KStar, SVM, MLP, and Decision tree and found that KStar gives some better accuracy than other algorithm which is 87.50%. Gadaleta et al. [23] presented an authentication framework which is IDNet based on user's way of walking. They detect human gait using accelerometer and gyroscope data placed on user's pocket. They used support vector machine to classify the cycle of walking and then used convolutional neural network to gait recognition. Karvekar [24] has done an impressive work which is fatigue detection using gait analysis. They used SVM algorithm and achieved good accuracy. They [25] studied human walking to detect Parkinson's diseases from 20 step of walking test. They used inertial sensors of a smartphone and applied machine learning approach. They have applied KNN, and Naive Bayes Algorithms. To identify different types of human gaits, a machine-learning-based system using a wearable accelerometer and an enhanced feature extraction algorithm is proposed [26]. They have applied KNN, SVM, Naïve Bayes, Decision tree and DA algorithm. They achieved good accuracy for short steps but for high steps and normal steps accuracy wasn't good. They used tri-axial accelerometer placed on leg to carried out the study. Nickel [27] proposed a biometric authentication on devices using gait analysis with Hidden Markov Model. They used phone's accelerometer for the data. Hayama et al. [28] studied a single accelerator sensor placing to a human waist to classify walking posture. They applied deep learning to detect walking posture and achieved 98.03% accuracy. Also, they applied SVM but find it difficult for SVM to classify. They applied single accelerator sensor whereas we applied accelerometer and gyroscope which is easily accessible and cost convenient in comparison to heuristic gait analysis. We used tri-axial accelerometer and gyroscope to detect abnormal gait. Similar works for human activity recognition were addressed in [31], where use of machine learning, hence deep learning were performed.

Different experiments are carried out with different techniques and method to analysis walking and posture. A lots of machine learning and deep learning approach have been found but only a few found working well. It is to be mentioned that all the work mentioned above, all are complex to perform. Nowadays smartphone is easily accessible everywhere so detecting abnormal gait is cheap and convenient with the proposed study.

Machine learning, Deep learning algorithms are intensively used in healthcare field. Recent days they are used in detecting breast cancer, tumor, human activity recognition, disease prediction, pattern of disease recognition etc. [29,30]. In most of the significant problems machine learning is used. That's why we applied several machine learning to detect abnormal gait. We have applied "K nearest neighbor (KNN), Logistic Regression (LR), Naïve Bayes (NB), Support vector machine (SVM), and two-dimensional convolutional Neural network (2D CNN)". A comparative analysis and summary of different approaches in detecting walking and different posture shown below in Table 1.

**Table 1.** Comparative analysis of different techniques used to detect walk along with different activities

| Publication | Sensor | Algorithms | Best Accuracy |
| --- | --- | --- | --- |
| Luo et al. [13] | IMU | Their proposed algorithm | 95.74% |
| Hsu et al. [17] | Inertial Sensor | NWFE | 90.5% |
| Arif et al. [18] | Acceleration Sensor | KNN, BPNN, Rotation Forest | 98% |
| B. Almaslukh et al. [21] | Accelerometer & Gyroscope | SAE | 97.5% |
| H. Chan et al. [22] | Accelerometer | KStar, SVM, MLP. LR | 87.5% |
| H. Hashiguchi et al. [25] | Inertial Sensor | KNN, Naïve Bayes | 75.3 % |
| Hayama et al. [28] | Accelerator | LSTM | 98% |

## 3 Proposed methodology

### 3.1 Research subject and instrument

In the current world smart device like smartphone and smartwatches are easily accessible device. People used to use or worn these devices now a days. There are some sensors integrated which is common across all those devices. Accelerometer and gyroscope are two of them. These sensors are being extensively used for different purposed. Accelerometer is used for measure acceleration and gyroscope is used for sense angular velocity. In this study, we have decided to use those sensors integrated along with smartphone to detect abnormal gait. We have classified normal and abnormal gait in our study. We used android smartphone which is running on Android 11 operating system and smartphone used is from vendor "Samsung" and the model is "Samsung Galaxy S10 Plus". The smartphone is placed on the front pocket of the subjects to record the activities. And a paid android app called "Sensor Data" is used to record activities.

### 3.2 Data collection process

Sensors creates numerous data at a time. So, collecting these data is a task of challenge. Firstly, we have managed twenty-three volunteers. Among of them fourteen people have normal gait and nine people have abnormal gait. The application used is set to 50hz sampling rate so that it can generate 50 samples per second. The smartphone with application running was placed on the subject's front pocket and we generate 1 minutes 6 seconds of data. We have included 1 minutes of data in the dataset. And cut the 6 seconds as first 3 seconds and last 3 seconds as transition time. Accelerometer and gyroscope are a tri-axial sensor. So, they give data of three axis that is X, Y, and Z. Thus, we have 7 columns. 6 columns generated from sensors and one column is for labeling the data whether it is normal gait or abnormal gate. We named the three columns of accelerometer as "ac_x, ac_y, ac_z" and three columns of gyroscope as "gy_x, gy_y, gy_z". Sensor data app generates the data into text files. We then converted the data into csv file. Then we merged all the csv into a comma separated value (CSV) file. Then, we removed the unnecessary strings and filled the fields containing null values. We preprocessed the raw data so that the algorithm can be applied. We had total 23 subjects and each second 50 row is generated. So, we have 69000 rows in our dataset. In 69000 samples, 42000 samples are generated from people with normal gait and rest 27000 samples are generated from people who had difficulty in walking. These 69000 rows then preprocessed.

### 3.3 Data preprocessing

After data collection we have processed our data. In this step we have done feature engineering. Firstly, we removed the null values by median methods. Also, noise presents in the dataset. We applied butterfly method to remove the noise. Out of 7 columns we have one row that has categorical data that is label column which consists of Label as normal or abnormal. We applied label encoding to that column to convert the categorical data into numerical data. Then we separated the data into two parts. One part consist only the input and other set consist of only output. We used K-Fold cross validation to train the model. We split out dataset with ratio of 4:1 for training and test data with distribution of 80–20% of data.

After that, we implemented feature scaling. We applied feature scaling to scale the data into a specific range. Standard Scaler and Robust scaler is used to scale the data. Also, scalar fitting and transformation is applied on the training dataset but only transformation is used in test dataset to prevent data leakage. For that test seen remains unseen to the model.

Though, the type of our dataset is time series we used sliding windows to specify the window size. We tried and observed different window size and their performance. We found that the bigger the window size is providing promising results. But the too big window side behind the reason of model overfitting. As a result, we decided to use 4 seconds as our window size. The hop size also known as stride size is also defined. So according to stride size in every iteration window size is moving forward with a size of 200 row and 6 columns per step. To reduce overlapping large hop size is used and it ensures the diversity.

To get 3D data we used sliding window. As mentioned above, we got 50 samples in each second. So, for a window of 4 second, we got 200 samples and 6 columns of data. This is the best fitted shape for neural networks thus we applied this dataset for applying 2D CNN algorithm.

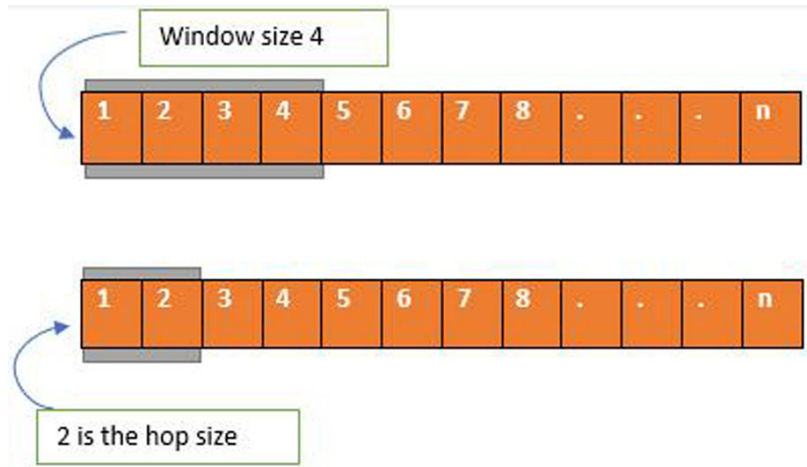

**Fig. 1.** Sliding window mechanism used in our dataset

This data can't pass through the traditional machine learning algorithm. For applying machine learning algorithm, we flattened the data. We have 200×06 shape in our dataset. After flattering the shape of the dataset is 1200 columns per activity. There are 1200 columns after flattening the data thus it can lead to model overfitting. So Principal Component Analysis (PCA) have performed to reduce the dimensionality.

After doing all these steps still there had been some problems during applying deep learning models. So, we reshape the whole dataset according to the algorithm requirement.

## 4 Results and discussion

There has been multiple work to recognize human gait and different postures and activities. But as I mentioned earlier all these works need lab environment and required assistance. Besides there most study found costly. In this study we introduced a way to detect abnormal walking and walking with a cost-effective way with smartphone's sensor. And we found that we can detect abnormal walking and normal walking with a good accuracy. We have applied KNN, Logistic Regression, Naive Bayes, Support Vector Machine and 2D CNN. In Table 2, we show the performance of these algorithms with their accuracy.

**Table 2.** Algorithms implemented with f1 score and accuracy

| Algorithms | F1 Score for Normal Walking | F1 Score for Abnormal Walking | Accuracy |
|---|---|---|---|
| KNN | 0.93 | 0.96 | 95% |
| Logistic Regression | 0.54 | 0.78 | 70% |
| Naïve Bayes | 0.75 | 0.88 | 84% |
| SVM | 0.94 | 0.97 | 96% |
| 2D CNN | 0.97 | 0.98 | 98% |

Among these algorithms implement Logistic regression perform worst. It has achieved accuracy of 70% with F1 score 0.54 and 0.78. Naïve Bayes performed much better than logistic regression. It has achieved 84% with accuracy of f1 score 0.75 and 0.88. KNN and SVM both performed well. They achieved accuracy of 95% and 96%. According to f1 score SVM performed slightly better than KNN. At least we applied deep learning algorithm which is 2D CNN. It gives us the optimum performance for our study. We got an accuracy of 98% with f1 score for normal walking is 0.97 ad f1 score for abnormal walking is 0.98 which is pretty good. Precision is achieved of normal walking is 0.96 and for abnormal walk is 0.98.

## 5 Conclusion

This work can be done in different techniques but as of now this is the most convenient and cost friendly techniques for detecting abnormal gait. There were some works done but these need lab environment and labor intensive. Hayama et al. [28] have achieved good accuracy of 98% with LSTM algorithm but for detecting walk this study also need special assistance or device. We have achieved 98% accuracy with 2D CNN algorithm with built in sensor of smartphone. It is cost effective and easy to accessible. Thus, it is easily can be used in healthcare or IoT based smart city apps or detect Parkinson's disease detection.

## 7 Authors

**Md Shahriar Tasjid** is from the Department of Computer Science and Engineering of Daffodil International University (DIU). He is a potential researcher in the field of machine learning, deep learning, wearable sensors, and artificial intelligence in healthcare.

**Ahmed Al Marouf** is currently working as a senior lecturer at the Department of Computer Science and Engineering of Daffodil International University (DIU). He is currently pursuing his Ph.D. in Computer Science at the Department of Computer Science of University of Calgary, Alberta, Canada. He is a potential researcher in the field of computational intelligence, mobile technologies, educational data mining, health informatics, and artificial intelligence in health science. E-mails: marouf.cse@diu.edu.bd; ahmedal.marouf@ucalgary.ca